\documentclass[10pt,twocolumn]{article}

\usepackage{multicol}
\usepackage[utf8]{inputenc}
\usepackage[english]{babel}
\usepackage{amsmath, amsthm, amsfonts, amssymb, slashed, url}
\usepackage{graphicx}
\usepackage{color}
\usepackage[margin=1.8cm]{geometry} 
\usepackage{authblk} 

\title{\bf The gauge freedom in the Aharonov-Casher theorem: \\ The problem in two and one dimensions }

\author{Lucas Sourrouille}
\affil{UNRN-CONICET, Anasagasti 1463 (8400), Bariloche, Rio Negro, Argentina \\ \texttt{sourrou@df.uba.ar}}

\date{\today}

\begin{document}

\twocolumn[
  \begin{@twocolumnfalse}
    \maketitle
    \begin{abstract}
     In this note, we investigate the role of gauge freedom in the Aharonov-Casher theorem in one and two dimensions. In particular, we analyze the asymptotic behavior of the gauge freedom. We show that in two-dimensional space the gauge field is uniquely determined, whereas in one dimensional space a residual gauge freedom allows for an infinite family of equivalent gauges. This freedom is determined by a constant k. However, the number $k$ must be restricted in order to guarantee the existence of a normalizable zero mode. This restriction is determined by the inequality $|k| < \frac{1}{2} \int dx B(x)$, where $B(x)$ is a scalar field localized in a finite region of space. We show that this condition is equivalent to the gauge field taking opposite-sign values  at $+\infty$ and $-\infty$. Finally, we illustrate these results with an explicit example.   
   \end{abstract}
    \vspace{0.5cm}
    \noindent \small{\textbf{Keywords:} Massless Dirac fermions, Zero energy modes, Aharonov-Casher theorem}
    \vspace{1cm}
  \end{@twocolumnfalse}
]

\vspace{1cm}
\section{Introduction}
Zero-energy states in quantum mechanics were first studied by Darwin \cite{D},
Fock \cite{F} and Landau \cite{L}. They considered a charged spin-1/2 particle interacting with a magnetic field in two-dimensional space.
The results of these studies showed that the Pauli equation possesses quantized energy levels of energy, where the presence of solutions with zero energy becomes natural.
For case of massless Dirac particles interacting with a magnetic field in two dimensions, Aharonov and Casher \cite{ahronov} were able to demonstrate 
the existence of zero-energy modes, which are polarized in the direction of the spin. In addition, the degeneracy of these zero mode was calculated, under the assumption that the magnetic ﬂux localized in a restricted region. They found that the number of states with zero energy for a one spin direction is finite and determined by magnetic flux.  
\\
The results of this theorem become important in the context of the study of electron states in graphene attached to a magnetic field, since this ensures that the zero mode are topologically
protected, which means that they are robust with respect to inhomogeneities in the magnetic
ﬁeld \cite{kats,kats1}. This theorem is particularly important in the context of quantum Hall effect in graphene \cite{k1,k2} 
\vspace{0.3cm}

In this note, we deal with the asymptotic behavior of the gauge freedom in the context of the Aharonov-Casher theorem.
We begin by reviewing the results of Aharonov and
Casher for the two-dimensional case. In two dimensions, the zero energy states are constructed in terms of a scalar potential $\lambda(x,y)$, from which the gauge fields and the magnetic field can be derived \cite{ahronov,sou}. With this in mind, we show that the normalizability requirement of the wave function imposes the condition that the gauge field be uniquely determined. In addition, we explore the one dimensional case. In that dimension, we show that the normalizability requirement of the wave function entails a one-parameter family of gauge-equivalent scalar potentials. However as we will see, the gauge freedom, determined by a constant $k$, is restricted by the inequality $|k| < \frac{1}{2} \int dx B(x)$, where $B(x)$ is a scalar field localized in a restricted region of the space. This conditon is equivalent to the gauge field taking opposite-sign values  at $+\infty$ and $-\infty$. We conclude this note by illustrating these situations with an example.

\section{The Aharonov-Casher theorem}
\label{5v}
We begin by considering a two-dimensional Dirac-Weyl model. The Hamiltonian is described by
\begin{equation}
H= \sigma^i p_i = (\sigma^1 p_1 +  \sigma^2 p_2)\;,
\label{}
\end{equation}
where, the $\sigma^i$ $(i =1,2)$
are the Pauli matrices,
\begin{eqnarray}
\sigma^1 =\left( \begin{array}{cc}
0 & 1 \\
1 & 0 \end{array} \right)
\,,
\;\;\;\;\;\
\sigma^2 =\left( \begin{array}{cc}
0 & -i \\
i & 0 \end{array} \right)
\end{eqnarray}
and $p_i =-i\partial_i$ is the two-dimensional momentum operator. Therefore we can write the massless Dirac-Weyl equation as
\begin{equation}
\sigma^i p_i \Phi (x, y, t) = i\partial_t \Phi (x, y, t)
\label{eq1}
\end{equation}
where, $\Phi (x, y, t)$ is two-component spinor
\begin{equation}
\Phi=(\phi_a,\phi_b)^T
\label{}
\end{equation}
In this note we will be interested on stationary states, so that we propose a
solution of the form
\begin{eqnarray}
\Phi (x, y, t) = e^{-iEt} \Psi (x, y)\;,
\label{}
\end{eqnarray}
Thus, we arrive to 
\begin{equation}
\sigma^i p_i \Psi (x, y) = E \Psi (x, y)
\label{1dw}
\end{equation}
When the charged particle interacts with 
a magnetic field along the $z$ direction, the momentum operator $p_i$ must be changed by the covariant
derivative; $D_{i}= -i\partial_{i} +A_{i}$ $(i =x,y)$, where $A_{i}$ are components of the vector potential. In terms of the vector potential the magnetic field may be read as
\begin{equation}
B=\partial_x A_y -\partial_y A_x
\label{mag}
\end{equation}
Then, instead of the equation (\ref{1dw}) we have,
\begin{equation}
\sigma^i D_i \Psi (x, y) = E \Psi (x, y)
\label{2dw}
\end{equation}
We can develop this equation and arrive to ,
\begin{eqnarray}
\Big[(-i\partial_x  -\partial_y ) + (A_x -i A_y) \Big]\psi_b = E \psi_a
\label{eqm1}
\end{eqnarray}
\begin{eqnarray}
\Big[(-i\partial_x  + \partial_y )+ (A_x +i A_y) \Big]\psi_a = E \psi_b
\label{eqm2}
\end{eqnarray}
where $\psi_a$ and $\psi_b$ are the components of the spinor $\Psi$.
\\
We can construct explicitly the solutions for zero energy. For this we will follow the work of Aharonov and Casher \cite{ahronov}. 
\\
First, we assume that the vector potential is divergenceless. 
Thus, we introduce a scalar potential $\lambda(x, y)$ such that,
\begin{eqnarray}
A_x = -\partial_y \lambda
\,,
\;\;\;\;\;\
A_y = \partial_x \lambda
\label{gau}
\end{eqnarray}
\begin{eqnarray}
B = \partial_x^2 \lambda + \partial_y^2 \lambda
\label{div}
\end{eqnarray}
In this way, the  equations (\ref{eqm1}) and (\ref{eqm2}) for zero energy solutions may be rewritten as 
\begin{eqnarray}
\Big[(-i\partial_x  -\partial_y ) + (-\partial_y \lambda -i \partial_x \lambda) \Big]\psi_b = 0
\label{eq1}
\end{eqnarray}
\begin{eqnarray}
\Big[(-i\partial_x  + \partial_y )+ (-\partial_y \lambda +i \partial_x \lambda) \Big]\psi_a = 0
\label{eq2}
\end{eqnarray}
So, we can propose as a solutions of equation (\ref{eq1}) and (\ref{eq2}) the following expressions
\begin{eqnarray}
\psi_b = f_b e^{-\lambda}
\label{pb}
\end{eqnarray}
\begin{eqnarray}
\psi_a = f_a e^{\lambda}
\label{pa}
\end{eqnarray}
By substitution of (\ref{pb}) and (\ref{pa}) in (\ref{eq1}) and (\ref{eq2}) respectively, we arrive to  
\begin{eqnarray}
[-i\partial_x -\partial_y] f_b=0
\label{fb}
\end{eqnarray}
and 
\begin{eqnarray}
[-i\partial_x +\partial_y] f_a=0
\label{fa}
\end{eqnarray}
This indicates to us that $f_b$ and $f_a$ are complex-conjugated analytic entire functions of the complex variable $z = ix + y$.
\\
Follow Ref.\cite{ahronov} we see that 
equation (\ref{div}) has a solution
\begin{eqnarray}
\lambda_{k_1 k_2} ({\bf r}) = \int  d {\bf r}'G({\bf r}, {\bf r}') B({\bf r}') + k_1 x + k_2 y 
\label{lambda}
\end{eqnarray}
where
\begin{eqnarray}
G({\bf r}, {\bf r}')= \frac{1}{2\pi}\ln \Big(\frac{|{\bf r} - {\bf r}'|}{r_0}\Big)
\end{eqnarray}
is the Green function of the Laplace operator in two dimensions. We denote by $r_0$ an arbitrary constant. 
Also $k_1$ and $k_2$ are arbitrary real numbers. 
Here, it is important to emphasize that Aharonov and Casher omit, in their article, the linear terms, $k_1 x + k_2 y$ in their solution. However, as we will see $k_1$ and $k_2$ must be zero in two-dimensional space, so that scalar potential $\lambda$ is uniquely determined.
We will see in the next section that the linear terms play an important role in the construction of $\lambda$ in one dimensional space, so that there exist an infinite family of scalar potentials. This is the main goal of this paper. 
\\
Since the magnetic flux is localized in 
a restricted region, the integral in formula (\ref{lambda}) only makes sense for restricted values of $r'$, so that in the limit $r \to \infty$ we can make the approximation 
where
\begin{eqnarray}
G({\bf r}, {\bf r}') \approx
G({\bf r})= \frac{1}{2\pi}\ln \Big(\frac{{\bf r}}{r_0}\Big)
\end{eqnarray}
Thus, at $r \to \infty$  the formula (\ref{lambda}) may be rewritten as 
\begin{eqnarray}
\lambda_{k_1 k_2} ({\bf r}) = \frac{\Theta}{2\pi} \ln \Big(\frac{r}{r_0}\Big) + k_1 x + k_2 y 
\label{}
\end{eqnarray}
where we denote $\Theta$ as the magnetic ﬂux.
The wave functions  (\ref{pb}) and (\ref{pa}) then may be written as
\begin{eqnarray}
\psi_{a,b} = f_{a,b} \Big(\frac{r}{r_0}\Big)^{\frac{\gamma \Phi}{2\pi}} e^{\gamma(k_1 x + k_2 y)}
\label{sol}
\end{eqnarray}
where $\gamma =1$ and $-1$ for $\psi_a$ and $\psi_b$ respectively.  Since the entire function $f(z)$ cannot go to zero in all
directions at infinity, $\psi_{a,b}$ can be normalizable only assuming that $k_1 = k_2 =0$ and $\gamma \Phi < 0$. In other words this means that  zero-energy states can
exist only for one spin direction, depending on the sign of the total magnetic flux. Furthermore, the fact that normalization imposes the condition $k_1 = k_2 = 0$ implies that the scalar potential is uniquely determined.In other words, the gauge fields $A_x$ and $A_y$ are also uniquely determined.
\\
In order to analyze the degeneracy of the solutions of the equations (\ref{eq1}) and  (\ref{eq2}) we need to determine the form of the functions $f_b$  and $f_a$. These are dictated by equations (\ref{fb}) and (\ref{fa}) and we can check are polynomials of the form
\begin{eqnarray}
f_{b} =z^j
\end{eqnarray}
\begin{eqnarray}
f_{a} = (z^\dagger)^j
\end{eqnarray}
Thus, we can see from (\ref{sol}) that the solutions are square integrable only assuming that
\begin{eqnarray}
j < N
\end{eqnarray}
where $N$ is the integer part of $\frac{\Phi}{2\pi}$.
Then, the number of the zero energy modes for one spin projection is equal to $N$, where we count from $j=0$ to $j=N-1$.

\section{The Aharonov-Casher theorem on a line}

In this section we consider the dimensional reduction of the model to
$1+1$ dimensions by assuming that the fields do not
depend on one of the spatial coordinates, say  $y$.
\\
Renaming  $A_y$ as $A$, and suppressing dependence on the $y$ coordinate, the equations (\ref{eqm1}) and (\ref{eqm2}) can be written as
\begin{eqnarray}
\Big[-i\frac{d }{dx}    + (A_x -i A) \Big]\psi_b = E \psi_a
\label{eqm1.1}
\end{eqnarray}
\begin{eqnarray}
\Big[-i\frac{d }{dx}   + (A_x +i A) \Big]\psi_a = E \psi_b
\label{eqm2.2}
\end{eqnarray}
Following the method proposed in  Ref.\cite{JPR}-\cite{JPR3}, the gauge field $A_x$ may be eliminated from the equations (\ref{eqm1.1}) and (\ref{eqm2.2}) via a gauge transformation.
Indeed, after transforming the matter fields as

\begin{eqnarray}
\psi_b(x)\rightarrow e^{-i\alpha(x)} \psi_b(x)\,, \,\,\,\,\,
\psi(x)_a\rightarrow e^{-i\alpha(x)} \psi_a(x)
\end{eqnarray}

with

\begin{eqnarray}
\alpha(x)=\frac{1}{2}\int dz \epsilon(x-z) A_x(z)
\label{alpha}
\end{eqnarray}
Then, the equations (\ref{eqm1.1}) and (\ref{eqm2.2}) reduce to 
\begin{eqnarray}
\Big[-i\frac{d }{dx}   -i A \Big]\psi_b = E \psi_a
\label{eqm1.1:1}
\end{eqnarray}
\begin{eqnarray}
\Big[-i\frac{d }{dx}  +i A \Big]\psi_a = E \psi_b
\label{eqm2.2:2}
\end{eqnarray}
Here, we are interested on the study of zero energy solutions of these equations. We can proceed as in the previous section, i.e., we can introduce the scalar potential $\lambda(x)$ such that
\begin{eqnarray}
A(x) = \frac{d \lambda (x)}{dx}
\label{gau}
\end{eqnarray}
so that the second derivative (the equivalent to the magnetic field in one dimension) reads
\begin{eqnarray}
B(x)= \frac{d^2 \lambda(x))}{dx^2}
\label{B1}
\end{eqnarray}
Then, it is not difficult to find the solutions of the equations (\ref{eqm1.1:1}) and (\ref{eqm2.2:2}) for the energy zero case. Indeed, substituting 
\begin{eqnarray}
\psi_b = e^{-\lambda }
\label{b1}
\end{eqnarray}
in equation (\ref{eqm1.1:1}) and setting $E=0$ we check easily that this is a solution. Similarly we can check that 
\begin{eqnarray}
\psi_a = e^{\lambda }
\label{a1}
\end{eqnarray}
is a zero energy solution for (\ref{eqm2.2:2})
\\
The important point, here, is to analyze the the solution of the (\ref{B1}). In this case the solution is 
\begin{eqnarray}
\lambda (x) = \int  d {x}'G({x}, x') B({x}') + k x 
\label{escalar}
\end{eqnarray}
where
\begin{eqnarray}
G(x, {x}')= \frac{1}{2}|x-x'|
\end{eqnarray}
so that
\begin{eqnarray}
\frac{d^2 (\frac{1}{2}|x-x'|)}{dx^2} = \delta (x-x')
\end{eqnarray}
Again, following the Aharonov-Casher arguments, we assume that $Q = \int  d {x} B({x})$ is localized in a restricted region so that for $x \to \mp\infty$
\begin{eqnarray}
\lambda (x) = \frac{1}{2}|x| Q+ k x
\label{scp}
\end{eqnarray}
In terms of this expression we can analyze the solutions (\ref{b1}) and (\ref{a1}). 
\\
First, consider the case $x>0$. Then, the scalar potential (\ref{scp}) reads as 
\begin{eqnarray}
\lambda (x) = (\frac{1}{2} Q+ k) x
\label{scp1}
\end{eqnarray}
Second, we can observe that if $x<0$, the potential (\ref{scp}) becomes
\begin{eqnarray}
\lambda (x) = (-\frac{1}{2} Q+ k) x
\label{scp2}
\end{eqnarray}
Since, we are interested in the cases in which the solutions (\ref{b1}) and (\ref{a1}) are normalizable functions, we require
\begin{eqnarray}
\lim_{x \to \pm\infty}\lambda (x) = +\infty 
\label{lim1}
\end{eqnarray}
for $\psi_b$ and 
\begin{eqnarray}
\lim_{x \to \pm\infty}\lambda (x) = -\infty 
\label{lim2}
\end{eqnarray}
for $\psi_a$. 
\\
The condition (\ref{lim1}) establishes 
\begin{eqnarray}
\frac{1}{2} Q+ k >0 \,, \,\,\,\,\,
-\frac{1}{2} Q+ k <0
\label{asb}
\end{eqnarray}
which lead us 
\begin{eqnarray}
-\frac{1}{2} Q  <k < \frac{1}{2} Q \,, \,\,\,\,\,
Q>0
\label{c1}
\end{eqnarray}
whereas from the condition (\ref{lim2}) we have
\begin{eqnarray}
\frac{1}{2} Q+ k <0 \,, \,\,\,\,\,
-\frac{1}{2} Q+ k >0
\label{asb1}
\end{eqnarray}
This implies 
\begin{eqnarray}
\frac{1}{2} Q  <k < -\frac{1}{2} Q \,, \,\,\,\,\,
Q<0
\label{c2}
\end{eqnarray}
Thus, as in the two dimmensional Aharonov-Casher theorem, the zero-energy solutions can
exist only for one spin direction, depending on the sign of the integral $Q$, but also the equations (\ref{c1}) and (\ref{c2}) imply conditions on the parameter $k$. These conditions establish, in order to the zero-energy solutions can
exist, that the parameter $k$ must be bounded as follows 
\begin{eqnarray}
|k| < \frac{1}{2} Q 
\label{|k|}
\end{eqnarray}
This condition indicates that the one-dimensional problem admits a family of gauge-equivalent scalar potentials $A_k(x)$.. Although the parameter k does not label physically distinct zero modes, it controls the asymptotic behavior of the gauge field and therefore determines whether the zero-energy solution is normalizable. This stands in stark contrast to the two-dimensional case. Furthermore, condition (\ref{|k|}) is equivalent to demanding that the field $A(x)$ takes opposite-sign values at $-\infty$ and $+\infty$. This can be easily seen since the field $A(x)$ has an asymptotic behavior dictated by formulas  (\ref{asb}) or  (\ref{asb1}), depending on the spin direction.
\\
We conclude this note with an example that illustrate the gauge freedom allowed by the formula (\ref{|k|}). Let us consider the localized scalar field

\begin{eqnarray}
B(x)=\frac{B_0}{a}\mathrm{sech}^2(x/a),
\qquad 
\label{Bx}
\end{eqnarray}
where $a$ and $B_0$ are real constants, and we assume that $B_0>0$.
The field $B(x)$ gives the total flux $Q$  

\begin{eqnarray}
Q=\int_{-\infty}^{\infty}B(x)dx=2B_0.
\end{eqnarray}
A corresponding solution of $\lambda''(x)=B(x)$ is
\begin{eqnarray}
\lambda_k(x)=aB_0\ln\cosh(x/a)+kx,
\label{lambdax}
\end{eqnarray}
so that

\begin{eqnarray}
A_k(x)=\lambda_k'(x)=B_0\tanh(x/a)+k
\label{Ax}
\end{eqnarray}
The zero-energy solution for the lower spinor component is therefore
\begin{eqnarray}
\psi_b(x)=
\exp\left[-aB_0\ln\cosh(x/a)-kx\right]
\end{eqnarray}
From this equation, the zero-energy solution can be written as
\begin{eqnarray}
\psi_b(x)
=
\frac{e^{-kx}}{\cosh^{aB_0}(x/a)}.
\label{bpsi}
\end{eqnarray}
Its asymptotic behavior determines whether the state belongs to
$L^2(\mathbb R)$. For $x\to+\infty$, one has
\begin{eqnarray}
\cosh(x/a)\sim \frac{1}{2}e^{x/a},
\end{eqnarray}
So that 
the wave function (\ref{bpsi}) has the following asymptotic behavior
\begin{eqnarray}
\psi_b(x)\sim \frac{e^{-kx}}{ \left(\frac{1}{2}e^{x/a}\right)^{aB_0}} =2^{aB_0}e^{-(B_0+k)x}
\end{eqnarray}
Thus, the wavefunction decays at $+\infty$ only if
\begin{eqnarray}
B_0+k>0.
\end{eqnarray}
On the other hand, for $x\to-\infty$
\begin{eqnarray}
\cosh(x/a)\sim \frac{1}{2}e^{-x/a},
\end{eqnarray}
Hence,
\begin{eqnarray}
\psi_b(x)\sim 2^{aB_0}e^{(B_0-k)x}.
\end{eqnarray}
Since $x\to-\infty$, this term decays only if
\begin{eqnarray}
B_0-k>0.
\end{eqnarray}
Consequently, the normalizability condition is
\begin{eqnarray}
-B_0<k<B_0.
\end{eqnarray}
Using $Q=2B_0$, this condition becomes
\begin{eqnarray}
|k|<\frac{Q}{2}.
\end{eqnarray}
\begin{figure*}[t]
\centering
\includegraphics[width=0.95\textwidth]{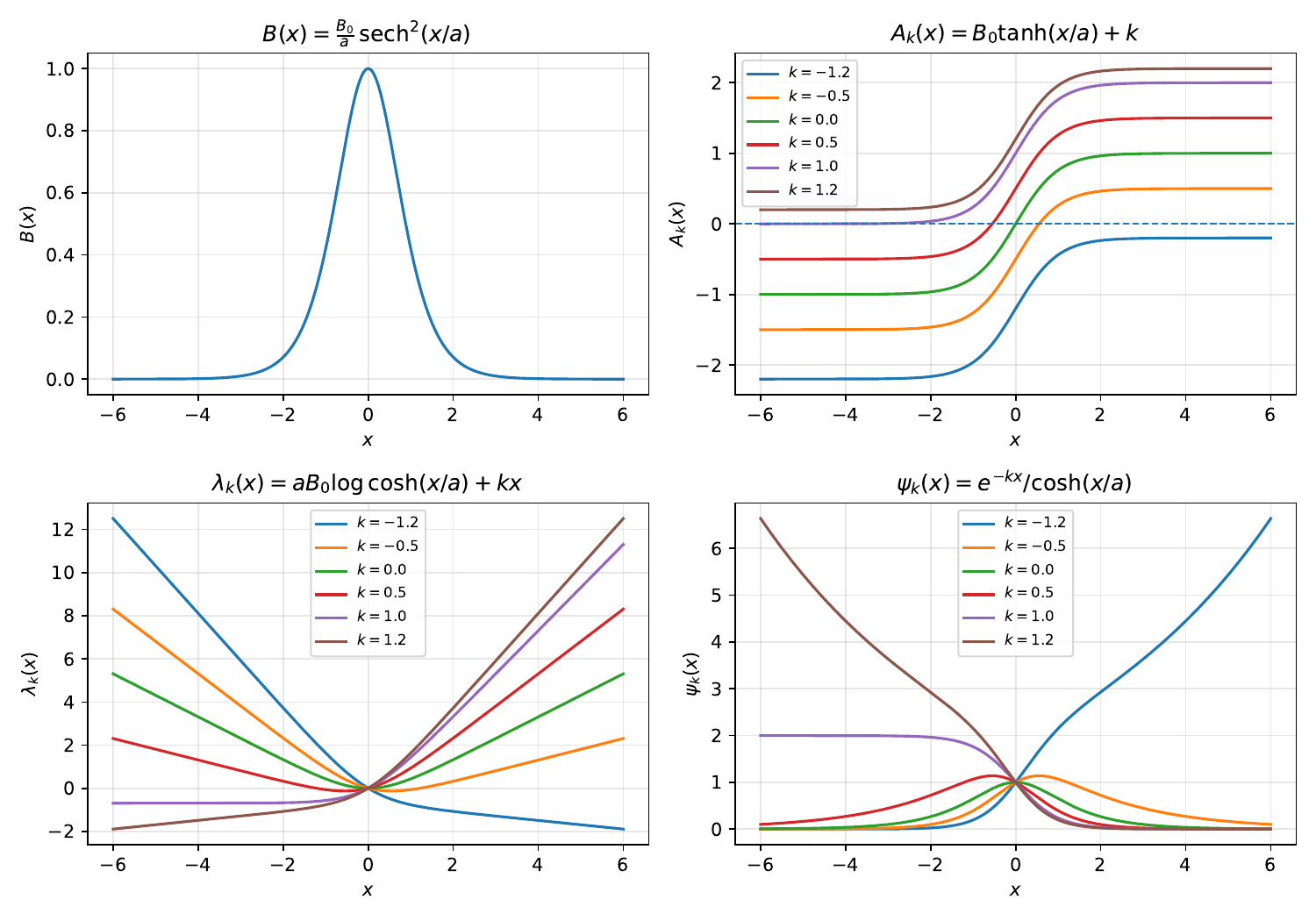}
\caption{
Illustration of the gauge freedom in the one-dimensional Aharonov–Casher problem.
Example with $B(x)=B_0\mathrm{sech}^2(x/a)$. For $a=B_0=1$,
the condition for a normalizable zero mode is $|k|<Q/2=1$. The example illustrates the role of the parameter $k$. The first three panels show the scalar field $B(x)$, the corresponding family of gauge-equivalent scalar and vector potentials, and the last panel displays the associated zero-energy wavefunctions. Only the values satisfying $|k|<Q/2$ produce square-integrable states.
}
\label{fig:AC-example}
\end{figure*}
½
This example illustrates the role of the linear homogeneous
term in the one-dimensional problem. The parameter $k$ does not
label independent degenerate zero modes; rather, it changes the
asymptotic behavior of the gauge field $A_k(x)$ and therefore
determines whether the formal zero-energy solution is square
integrable. In this sense, Eq.(\ref{|k|}) should be interpreted as a
normalizability condition on the admissible asymptotic gauges.
\\
Finally, Fig. (\ref{fig:AC-example})illustrates the profiles of the functions (\ref{Bx}) (\ref{lambdax}) (\ref{Ax}) and (\ref{bpsi}). In the profile of the function $\lambda_k(x)$ we see that the limit (\ref{lim1}) is satisfied for the values of $k$ restricted by $|k|<1$.  
Similarly, in the profile of the function $A_k(x)$, it is possible to see that for values of $k$ satisfying $|k|<1$, the condition $A_k(-\infty)A_k(+\infty)<0$ holds, which is the condition we had established earlier for the fields $A_k(x)$. Lastly, we have also plotted the wave function $\psi_k(x)$ for different values of the parameter $k$. It can be clearly seen that the square-integrable functions are precisely those satisfying the inequality $|k|<1$.

\section{Conclusion}

In this note, we have revisited the role of gauge freedom in the Aharonov--Casher construction of zero-energy modes. We first reviewed the original two-dimensional theorem and showed that the requirement of normalizability of the wave function uniquely determines the asymptotic behavior of the scalar potential $\lambda(x)$. Then, the requirement of normalizability forces the coefficients of the linear homogeneous solutions of the Poisson equation to vanish, so that the gauge field is uniquely determined.
\\
We then considered the one-dimensional reduction of the problem. In contrast with the two-dimensional case, the normalizability condition does not force the coefficient of the homogeneous linear contribution to vanish. Instead, it allows a finite interval of values of the parameter $k$. 
This difference originates from the different asymptotic constraints imposed 
by the normalizability condition on the zero-energy wavefunction in one dimension.
\\
Assuming that the scalar field $B(x)$ is localized in a finite region of space, we found that the asymptotic form of the scalar potential is 
\begin{eqnarray}
\lambda (x) = \frac{1}{2}|x| Q+ k x,
\label{scp}
\end{eqnarray}
so that the condition of the  normalizability of the wave function imposes a restriction on the parameter $k$
\begin{eqnarray}
|k|<\frac{1}{2}\int_{-\infty}^{\infty} B(x)dx
\end{eqnarray}
This condition is equivalent to demanding that the field $A(x)$ takes opposite sign values at -$\infty$ and +$\infty$
\\
Finally, we illustrated these results with an explicit example. The family of gauge-equivalent scalar and vector potentials provides a simple visualization of how the asymptotic gauge freedom controls the existence of normalizable zero-energy states.
We hope that this analysis provides a useful pedagogical complement to the original Aharonov--Casher theorem and clarifies the role played by the homogeneous solutions of the Poisson equation in one-dimensional reductions.

\vspace{0.6cm}

{\bf Acknowledgements}
\\
This work was supported by Universidad Nacional de Río Negro through Research Project PI UNRN 40-B-1286, and by CONICET.

\end{document}